\documentclass[sigconf,natbib=true, authorversion]{acmart}

\AtBeginDocument{%
  \providecommand\BibTeX{{%
    \normalfont B\kern-0.5em{\scshape i\kern-0.25em b}\kern-0.8em\TeX}}}

\copyrightyear{2023}
\acmYear{2023}
\setcopyright{acmlicensed}\acmConference[WWW '23 Companion]{Companion Proceedings of the ACM Web Conference 2023}{April 30-May 4, 2023}{Austin, TX, USA}
\acmBooktitle{Companion Proceedings of the ACM Web Conference 2023 (WWW '23 Companion), April 30-May 4, 2023, Austin, TX, USA}
\acmPrice{15.00}
\acmDOI{10.1145/3543873.3587654}
\acmISBN{978-1-4503-9419-2/23/04}

\usepackage{algorithm}
\usepackage[noend]{algpseudocode}
\usepackage[font=small,labelfont=bf]{caption}
\usepackage{balance} 
\usepackage{graphicx}
\usepackage{subfig}
\usepackage{mathrsfs}

\newcommand{\V}[0]{\mathcal{V}}
\newcommand{\E}[0]{\mathcal{E}}
\newcommand{\GG}[0]{\mathcal{G}}
\newcommand{\X}[0]{\mathcal{X}}
\newcommand{\CC}[0]{\mathscr{C}}
\newcommand{\M}[0]{\mathcal{M}}
\newcommand{\cvec}[0]{\mathbf{c}}

\newcommand{\secoder}[0]{\textit{Snapshot Encoder}}
\newcommand{\qencoder}[0]{\textit{Query Encoder}}
\newcommand{\bb}[0]{\mathbf{b}}
\newcommand{\W}[0]{\mathbf{W}}
\newcommand{\Dr}[0]{\mathrm{Dr}}

\newcommand{\head}[1]{\vspace{1.7mm}\noindent{{\bf #1.}}}

\newtheoremstyle{sig}
  {}
  {}
  {\itshape}
  {}
  {\scshape}
  {.}
  {.5em}
  {#1 #2\thmnote{\quad(#3)}}

\theoremstyle{sig}

\newtheorem{problem}{Problem}

\newtheorem{remark}{Remark}

\newcommand{\eat}[1]{}

\newcommand{\squishlist}{
 \begin{list}{$\bullet$}
  { \setlength{\itemsep}{0pt}
     \setlength{\parsep}{3pt}
     \setlength{\topsep}{3pt}
     \setlength{\partopsep}{0pt}
     \setlength{\leftmargin}{1.5em}
     \setlength{\labelwidth}{1em}
     \setlength{\labelsep}{0.5em} } }

\newcommand{\squishlisttwo}{
 \begin{list}{$\bullet$}
  { \setlength{\itemsep}{0pt}
    \setlength{\parsep}{0pt}
    \setlength{	opsep}{0pt}
    \setlength{\partopsep}{0pt}
    \setlength{\leftmargin}{2em}
    \setlength{\labelwidth}{1.5em}
    \setlength{\labelsep}{0.5em} } }

\newcommand{\squishend}{
  \end{list}  }
\newcommand{\model}{\textsc{CS-TGN}}
\begin{document}
\setlength{\textfloatsep}{5pt}

\title{\model: Community Search via Temporal Graph Neural Networks}


\author{Farnoosh Hashemi}  
\authornote{$\:$All authors contributed equally.}
\affiliation{%
  \institution{University of British Columbia}
  \city{Vancouver}
  \state{BC}
  \country{Canada}
}
\email{farsh@cs.ubc.ca}

\author{Ali Behrouz}
% \authornote{\relax}
\authornotemark[1]
\affiliation{%
  \institution{University of British Columbia}
  \city{Vancouver}
  \state{BC}
  \country{Canada}
}
\email{alibez@cs.ubc.ca}

\author{Milad Rezaei Hajidehi}
% \authornote{\relax}
\authornotemark[1]
\affiliation{%
  \institution{University of British Columbia}
  \city{Vancouver}
  \state{BC}
  \country{Canada}
}
\email{miladrzh@cs.ubc.ca}

\newcommand{\blue}[1]{\textcolor{blue}{#1}}

\begin{abstract}
  Searching for local communities is an important research challenge that allows for personalized community discovery and supports advanced data analysis in various complex networks, such as the World Wide Web, social networks, and brain networks. The evolution of these networks over time has motivated several recent studies to identify local communities in temporal networks. Given any query nodes, Community Search aims to find a densely connected subgraph containing query nodes. However, existing community search approaches in temporal networks have two main limitations: (1) they adopt pre-defined subgraph patterns to model communities, which cannot find communities that do not conform to these patterns in real-world networks, and (2) they only use the aggregation of disjoint structural information to measure quality, missing the dynamic of connections and temporal properties. In this paper, we propose a query-driven Temporal Graph Convolutional Network (\model) that can capture flexible community structures by learning from the ground-truth communities in a data-driven manner. \model{} first combines the local query-dependent structure and the global graph embedding in each snapshot of the network and then uses a \textsc{GRU} cell with contextual attention to learn the dynamics of interactions and update node embeddings over time. We demonstrate how this model can be used for interactive community search in an online setting, allowing users to evaluate the found communities and provide feedback. Experiments on real-world temporal graphs with ground-truth communities validate the superior quality of the solutions obtained and the efficiency of our model in both temporal and interactive static settings.
\end{abstract}


\begin{CCSXML}
<ccs2012>
   <concept>
       <concept_id>10010147.10010257.10010293.10010294</concept_id>
       <concept_desc>Computing methodologies~Neural networks</concept_desc>
       <concept_significance>300</concept_significance>
       </concept>
   <concept>
       <concept_id>10002950.10003624.10003633.10010917</concept_id>
       <concept_desc>Mathematics of computing~Graph algorithms</concept_desc>
       <concept_significance>300</concept_significance>
       </concept>
   <concept>
       <concept_id>10002951.10003260.10003282.10003292</concept_id>
       <concept_desc>Information systems~Social networks</concept_desc>
       <concept_significance>300</concept_significance>
       </concept>
 </ccs2012>
\end{CCSXML}



\maketitle

\vspace*{-1ex}
\section{Introduction}
Identifying communities is a fundamental problem in network science, where the aim is to partition a network into subgraphs that effectively capture dense groups of nodes with strong interconnections~\cite{Community_Detection_main}. With the increasing focus on personalized communities, the query-dependent community discovery problem, known as community search (CS)~\cite{k-core-community}, has become a popular area of research with a wide array of applications, including the classification of brain networks~\cite{FirmTruss}, modeling of social contagion~\cite{social_contagion}, content recommendation~\cite{Personal_content}, and team formation~\cite{TeamFormation}. Given a graph and a set of query nodes, the CS problem aims to find a cohesive and dense subgraph containing the query nodes~\cite{k-core-community}.  

\noindent
CS over static graphs has received significant research effort~\cite{k-core-community, clique-community, D-truss_community} while in many real-world complex networks such as web, social, communication, transportation, and brain networks, the interactions between objects are dynamic and subject to change over time. To this end, CS in temporal networks recently has attracted attention and several community models have been proposed~\cite{temporal-community, Adaptive, temporal-quasi, reliable, stable, temp1, truss-triangle}. However, these models suffer from three main limitations. \textbf{(1)} Existing methods mostly focus on how to efficiently update the community patterns (e.g., $k$-truss~\cite{truss-triangle}) and also assess the community quality using their aggregated structural cohesiveness at independent timestamps and ignore the structural dynamics of a community over time, missing the temporal properties.
 \textbf{(2)}   Structure inflexibility refers to the problem that they are based on pre-defined patterns or rules, e.g., \cite{temporal-community, temporal-quasi}. However, the structure of a community is flexible in nature, and it is nearly impossible to produce a high-quality community directly using the pre-defined rules. \textbf{(3)} These methods only consider the structural properties of the network, while in real-world applications, networks often come naturally endowed with attributes associated with their nodes, and these attributes can change over~time.

\noindent
Moreover, even in static graphs, existing methods face challenges when applied to real-life networks. Most existing community search approaches in static graphs use a progressive method to find the community~\cite{community_search_survey, D-truss_community, D-core_community_search}. However, based on each presented result for a given query set of nodes, users may need to adjust their query, select representative attributes, and modify the community size to achieve a desired high-quality result. Recently, Gao et al.~\cite{GNN_CS1} proposed an interactive GNN-based community search framework in static graphs, known as ICS-GNN, to address this issue. Nevertheless, the ICS-GNN framework requires the whole model to be re-trained for each query, which is time-consuming and limits its practical applications in real-world scenarios, especially in the case of online queries. Additionally, ICS-GNN only modifies the initial candidate set of nodes based on user input and lacks the ability to learn the dynamics of user queries to adjust the results accordingly.

\begin{figure*}
\centering
    % \hspace*{-3ex}
    \subfloat[][\centering Architecture of the \model{} model]{{\includegraphics[width=0.37\textwidth]{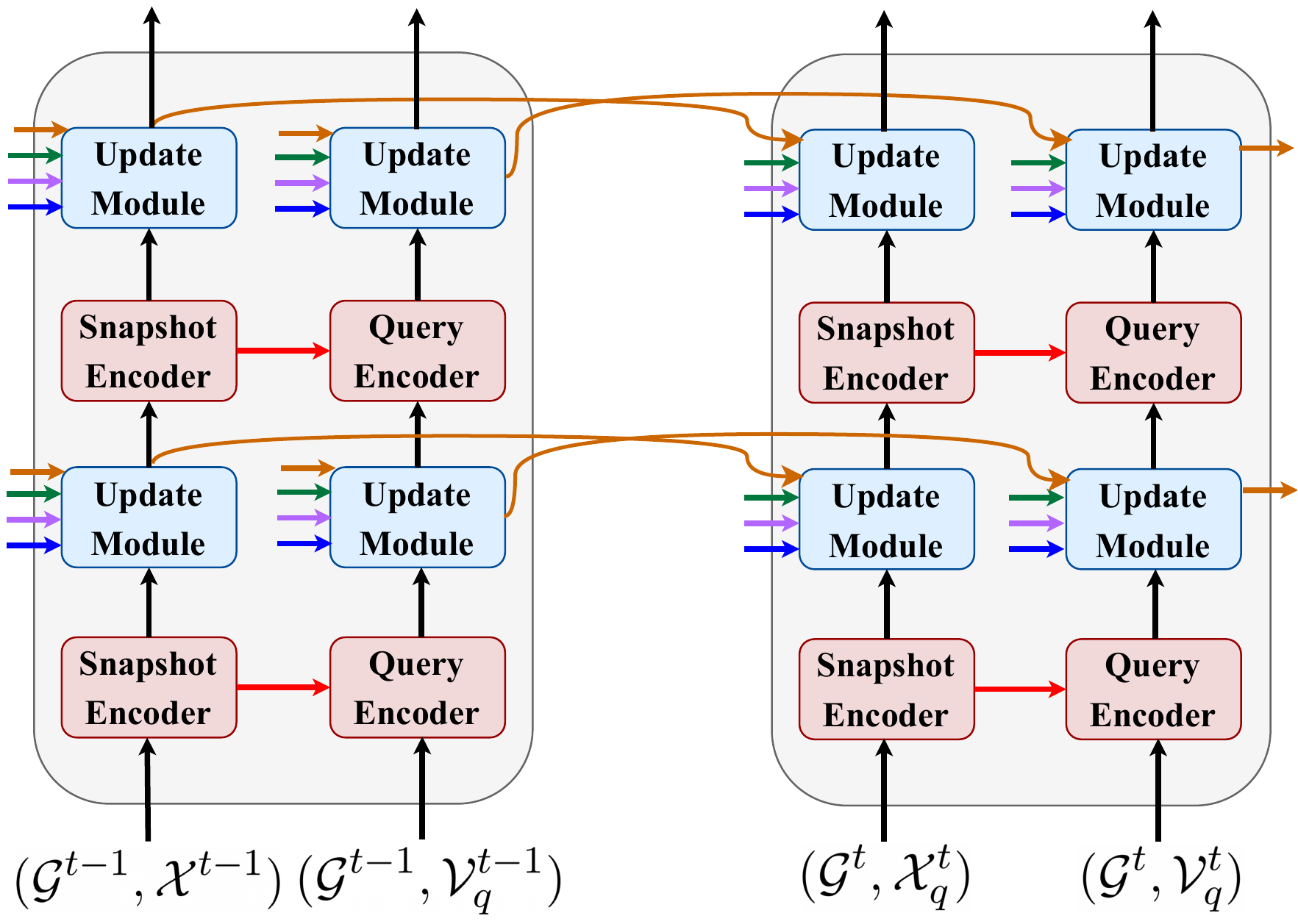} }}
    % \hspace*{2ex}
    \subfloat[][\centering Attention-based update module]{{\includegraphics[width=0.30\textwidth]{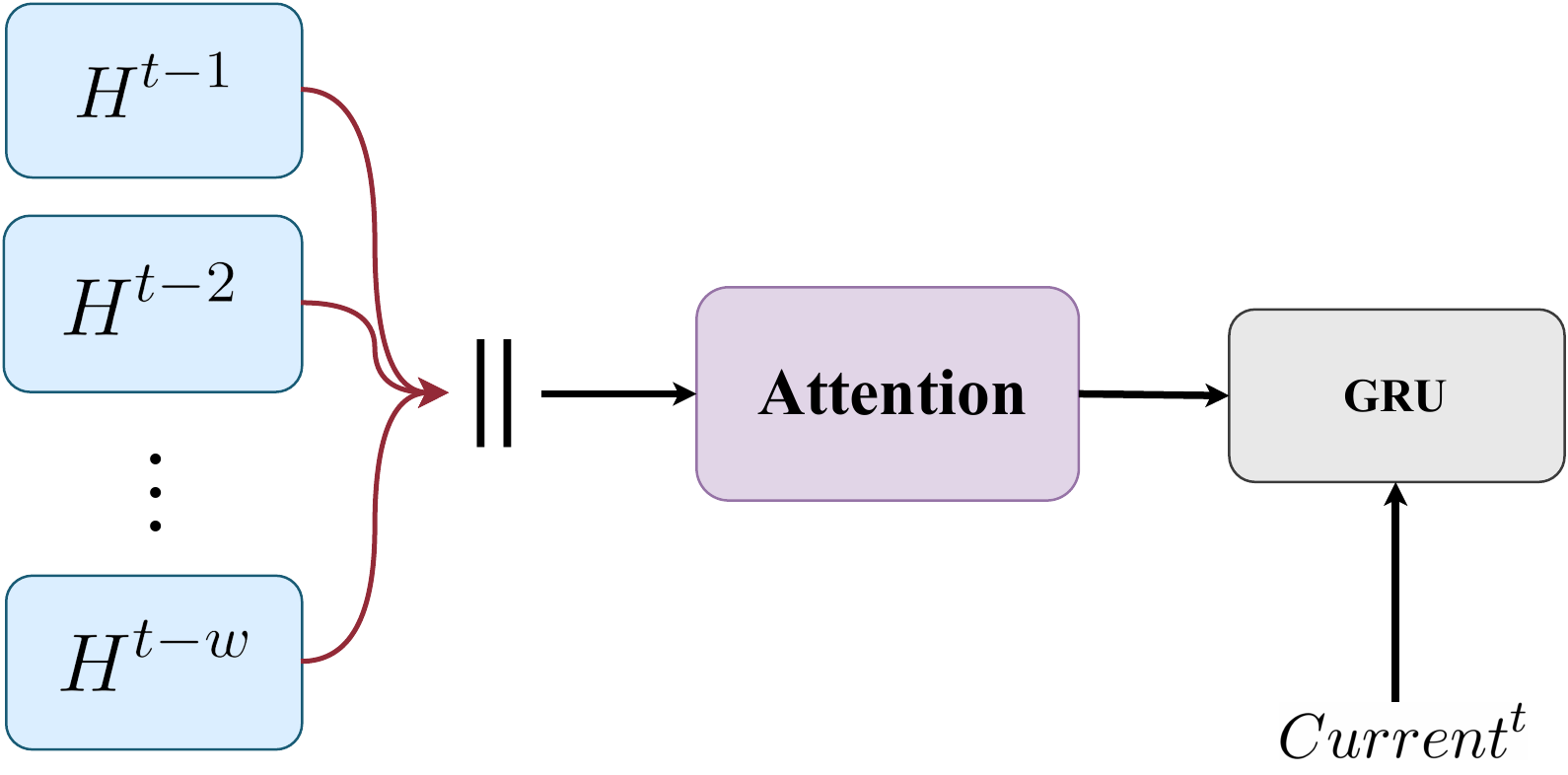} }}
    % \hspace*{2ex}
    \subfloat[][\centering Interactive CS via \model{} model]{{\includegraphics[width=0.33\textwidth]{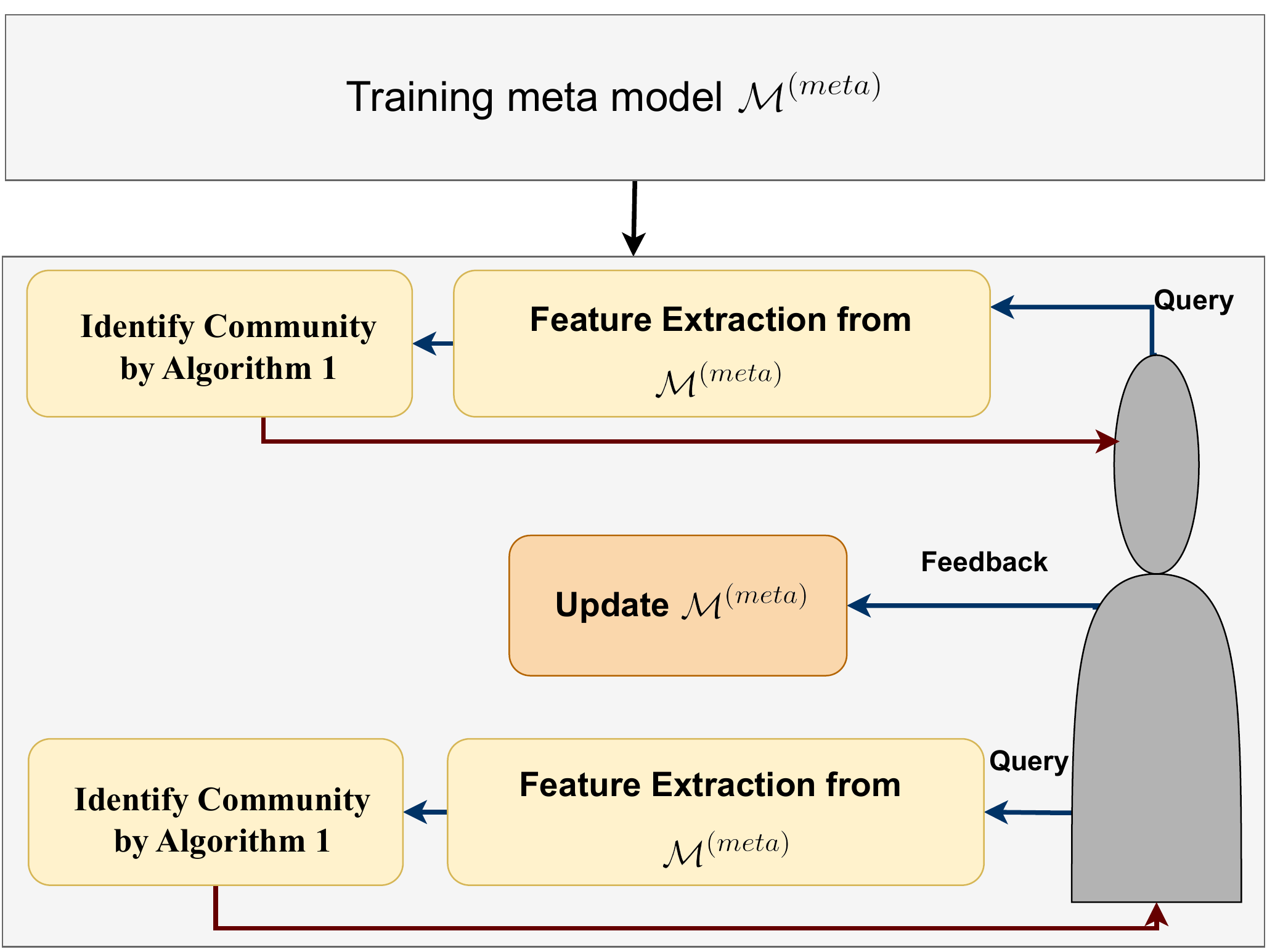} }}
    \vspace{-1ex}
    \caption{The design of \model{} framework.}
    \vspace{-2ex}
    \label{fig:model_framework}
\end{figure*}

\noindent
To mitigate the aforementioned limitations, we design a query-driven temporal graph convolutional neural network, called \model{}. To address structure inflexibility, we design two encoders: \textbf{(1)}~\textit{Query Encoder} to encode the structural information from query nodes and capture the local topology around the queries in each snapshot of the network, and \textbf{(2)}~\textit{Snapshot Encoder} to learn the query-independent node embeddings by combining the global structure and attributes of vertices in each snapshot of the network. To take advantage of node attributes, structural properties, and their dynamics in different snapshots of the network, \model{} employs Gated Recurrent Unit (GRU) cells~\cite{GRU} and extends the idea of hierarchical node states~\cite{roland} by using a contextual attention mechanism that can combine the hidden states for long-term patterns and the window information containing the short-term patterns. Next, we showcase the applicability of \model{} in interactive community search and propose a meta-learning framework. Within this framework, each query of the user is treated as an input of the \textit{Query Encoder} in a snapshot of the underlying network. We then consider the identification of communities based on user queries as distinct tasks. This design enables the model to \textbf{(1)} learn the dynamics of the network and better adapt to different user queries, and \textbf{(2)} learn the dynamic of user queries, resulting in better performance.

\noindent
To summarize, We make the following contributions:
\squishlist
\item We present \model, a query-driven temporal graph convolutional neural network that integrates the local query-dependent structure and global node embeddings at each timestamp. \model{} represents the local query-dependent and global node embeddings as hierarchical node states at different GCN layers and uses attention-based GRU cells to capture the network dynamics and update query-dependent and global embeddings over time. 
\item We demonstrate how \model{} can be utilized for interactive community search. Specifically, we approach interactive community search as a meta-learning problem over temporal networks, treating the network in each user-model interaction as one snapshot in \model{} framework. We then consider identifying communities based on queries from different users as distinct tasks.
\item By conducting extensive experiments on real-world temporal and static networks with ground-truth communities, we demonstrate that our method is capable of efficiently and effectively discovering communities in both online and interactive settings. 
\squishend

\section{Related Work}
\head{Community Search}
The concept of community search, which aims to find query-dependent communities in a graph, was first introduced by Sozio and Gionis \cite{k-core-community}. Since then, various community models have been proposed based on different pre-defined dense subgraphs~\cite{community_search_survey}, including $k$-core~\cite{k-core-community, community2}, $k$-truss~\cite{closest, TrussEquivalence}, quasi-clique \cite{clique-community}, $k$-plex~\cite{k-plex}, and densest subgraph~\cite{densest_community}. Recently, community search has also been explored in directed~\cite{community-directed, D-core_community_search}, weighted~\cite{weighted-truss}, geo-social~\cite{geo-social-community, road_social}, multilayer~\cite{FirmTruss, ML-LCD, CS-MLGCN}, multi-valued \cite{multi-valued}, and labeled~\cite{butterfly_core} graphs. Inspired by the success of Graph Neural Networks (GNNs), recently, several GNN-based approaches have been proposed for community search, such as \cite{GNN_CS, GNN_CS1, CS-MLGCN}. However, these models only focus on static networks and do not capture the dynamics of interactions and temporal properties.

\head{Community Search in Dynamic Networks}
CS in temporal networks recently has attracted attention and several community models have been proposed~\cite{temporal-community, Adaptive, temporal-quasi, reliable, stable, temp1}. Li et al.~\cite{temporal-community} have defined the persistent community search as the maximal $k$-core, where each vertex's cumulative degree satisfies the $k$-core requirement within a specified time interval. Qin et al.~\cite{stable} have proposed stable communities by first selecting centroid vertices with a certain number of neighbors having the desired similarity and a star-shaped structure of the centroid vertex and its neighbors existing frequently over a period of time, followed by clustering network vertices into stable groups based on the selected centroids. Similarly, Lin et al.~\cite{temporal-quasi} have defined frequency-based dense subgraphs that satisfy the quasi-clique structure with at least $\theta$ vertices, and each vertex has a degree more than a given threshold. Finally, Tang et al.~\cite{reliable} have introduced the Reliable CS problem based on $k$-core structure and the duration of interactions. These methods differ from our approach, as they are based on pre-defined patterns.

\head{Temporal Graph Learning}
Many approaches have been proposed in the literature to address the problem of learning from temporal networks~\cite{time-then-graph, dynamic_gnn1, dynamic_learning1, dynamic_learning2, dynamic_learning3, dynamic_learning4, dynamic_rnn2}. One group of methods u Recurrent Neural Networks (RNN) and replace the linear layer with a graph convolution layer~\cite{dynamic_rnn1, dynamic_rnn2, dynamic_rnn3}. Another group deploys a GNN as a feature encoder and a sequence model on top of the GNN to capture temporal properties~\cite{dynamic_gnn1, dynamic_gnn2, dynamic_gnn3}. However, these models have limitations in both their design and training strategies~\cite{roland}. To overcome these limitations, recent frameworks such as \textsc{ROLAND} and its variants~\cite{roland, anomuly} have been proposed to re-purpose static GNNs for dynamic graphs. However, these approaches only incorporate node embeddings from the previous snapshot in the GRU cell, failing to capture short-term patterns. In contrast, our proposed framework extends \textsc{ROLAND} by incorporating both long-term and short-term patterns using a contextual attention mechanism. Notably, our framework is designed for the CS problem and can capture query-dependent structural properties, which is not addressed by previous approaches.

\section{Problem Formulation}
We first precisely define temporal networks, and then we formalize the problem of community search in temporal graphs. Let $\GG = (\V, \E, \X) = \{\GG^t\}_{t = 1}^{T}$ denote a temporal network, where $\GG^t = (\V^t, \E^t, \X^t)$ represents the $t$-th snapshot of the network, $\V = \bigcup_{t = 1}^{T}\V^t$ is the set of nodes, $\E = \bigcup_{t = 1}^{T} \E^t$ is the set of edges, and $\X = \X^T \in \mathbb{R}^{|\V|\times f}$ is a matrix that encodes node attribute information for nodes in $\V$. Given attribute matrix $\X$, $\X^t_v$ represents the attribute set of vertex $v \in \V^t$ at timestamp $t$. We denote the set of vertices in the neighborhood of $u \in \V$ in $t$-th snapshot as $\mathcal{N}^t(u)$. Each edge $e = (u, v) \in \E$ is associated with a timestamp $\tau_e$, and each node $v \in \V$ is associated with a timestamp $\tau_v$.

\begin{problem}[Community Search in Temporal Networks]\label{problem:community_search}
Given a temporal network $\GG = \left\{ \GG^{1}, \dots, \GG^{t} \right\} = (\V, \E, \X)$, and a vertex query set $\V_q \subseteq \V$, the problem of Community Search in Temporal Networks (CST) is to find the query-dependent community $\CC_q \subseteq \V$ that is connected and has a cohesive structure. 
\end{problem}

\noindent
In this paper, we formulate the problem as a binary classification task. Given a set of query vertices $\V_q \subseteq \V$, we classify the nodes in $\V$ into: \textbf{1}: being a part of the community $\CC_q$, \textbf{0}: not being a part of it. We use a one-hot vector $\cvec^{\text{out}}_q \in \{0, 1\}^{|\V|}$ to represent the output community $\CC_q$ produced by model $\M$. Accordingly, if ${\cvec^{\text{out}}_{q_v}} = 1$, vertex $v$ is a part of the predicted community $\CC_q$ by~$\M$.

\section{\model{} Framework}
This section presents our proposed framework for Community Search (CS) in temporal networks. The framework consists of two main stages: offline training and online query, as illustrated in Figure~\ref{fig:model_framework}. In the offline training stage, we train a model denoted as $\M$ to predict the membership of each vertex to the corresponding community of query vertices. Specifically, $\M$ learns to capture the flexible community structures and the dynamic temporal properties of the network from ground-truth communities. In the online query stage, given a query vertex set $\V_q$, we utilize the trained model $\M$ to identify the corresponding community of $\V_q$.

\subsection{Architecture}
In our framework, given a timestamp $t$, we begin by utilizing GCN layers to encode both the graph structure and the features of the nodes present in the recent snapshot $\GG^t = (\V^t, \E^t, \X^t)$.  We then use GCN layers and the query vector $\cvec^t_q$ as the features of nodes to better capture the local query structure information. This part, called \qencoder, propagates from the query vertices to its surrounding nodes, allowing for query-centered structural propagation. We then use a contextual attention-based model with GRU cells to capture both short-term and long-term patterns of nodes and update the node embeddings over time. Accordingly, we incorporate the historical and temporal properties of the network. Finally, a feedforward neural network (FNN) is employed for classification. The architectures of \secoder{} and \qencoder{} are illustrated in Figure~\ref{fig:model_framework}(a).  Next, we explain each part in detail.

\head{Snapshot Encoder}
Given a timestamp $t$ and a snapshot of the network $\GG^t = (\V^t, \E^t, \X^t)$, \secoder{} captures the structural properties as well as attributes of vertices at timestamp $t$. To this end, it employs the layer-wise forward propagation of GCN with self-feature modeling as:
\begin{equation}\label{eq:propagatiion}
    {h_u^t}^{(\ell + 1)} \hspace{-0.5ex} = \Dr\left\{\sigma \left({h_u^t}^{(\ell)} {\W_{\text{s}}}^{(\ell + 1)}\hspace{-0.5ex} + \hspace{-2ex} \sum_{v \in N^t(u)} \hspace{-1ex} [\frac{{h_v^t}^{(\ell)}}{\sqrt{p^t_vp^t_u}} {\W}^{(\ell + 1)} \hspace{-0.5ex} + {\bb}^{(\ell + 1)}] \right)\right\},
\end{equation}
where at timestamp $t$, ${h^t_u}^{(\ell + 1)} \in \mathbb{R}^{d^{(\ell + 1)}_r}$ is the learned new features of node $u$ in the $(\ell + 1)$-th GCN layer, ${h^t_u}^{(\ell)} \in \mathbb{R}^{d^{(\ell)}_r}$ is the hidden feature of $u$ in $\ell$-th GCN layer, and ${\W_{\text{s}}}^{(\ell + 1)}, {\W}^{(\ell + 1)} \in \mathbb{R}^{d^{(\ell)} \times d^{(\ell + 1)}}$, and ${\bb}^{(\ell + 1)} \in \mathbb{R}^{d^{(\ell + 1)}}$ are trainable weights. $\sigma(.)$ is a nonlinearity, e.g., ReLU, and $\Dr(.)$ is the dropout method~\cite{dropout} to avoid overfitting. Given a vertex $u \in \V$, $p^t_u$ denotes the degree of node $u$ plus one at timestamp $t$, i.e., $p^t_u = |\mathcal{N}^t(u)| + 1$. The input feature of node $u$ in the first layer, ${h^r_u}^{(0)} \in \mathbb{R}^d$, is the normalized feature vector $\X^t_u$.

\head{Query Encoder}
At the time $t$, we transform a set of query vertices, denoted as $\V^t_q$, into a one-hot vector representation, called $\cvec^t_q$. That is, if a vertex $u$ is in the query set, then the value of ${\cvec^t_q}_u$ is set to 1, otherwise it is 0. Next, we apply the propagation function defined in Equation~\ref{eq:propagatiion} to each vertex $u\in\V^t$. We denote the query-dependent hidden features of vertex $u$ at $\ell$-th layer as ${{h^t_Q}_u}^{(\ell)}$, and use trainable query-dependent weights denoted by ${\W_{Q}}^{(\ell + 1)}{\text{s}}$, ${\W_Q}^{(\ell + 1)}$, and ${\bb_Q}^{(\ell + 1)}$. Unlike \secoder, where the input feature for vertex $u$ in the first layer is the node feature vector, in \qencoder, we use the one-hot query vector ${\cvec^t_q}_u$ as the input for the first GCN layer. As shown in Figure~\ref{fig:model_framework}(a), we combine the output of each layer in both \secoder{} and \qencoder{} and use it as the input for the next layer of \qencoder. This enables us to provide a stable and reliable knowledge of the graph's structure and node features that is independent of the query.

\head{Attention-based Update Module}
In order to capture the short-term patterns of nodes, inspired by~\cite{Attention-context}, we employ a contextual attention-based mechanism proposed by~\cite{cui2019hierarchical}. Specifically, for a given local window size $w$, we construct the short state of the window as follows:
\begin{align}
    &{C^{t}_u}^{(\ell)} = \left[ h_u^{{t - w}^{(\ell)}}, \dots, h_u^{{t - 1}^{(\ell)}} \right],\\
    &{E^{t}_u}^{(\ell)} = \mathrm{softmax}\left(\mathbf{r}^{t^{(\ell})} \tanh \left( \mathbf{Q}^{t^{(\ell)}}  \left( {C^{t}_u}^{(\ell)} \right)^T\right)\right),\\
    &\mathrm{short}_u^{t^{(\ell + 1)}} = \left({E^{t}_u}^{(\ell)} {C^{t}_u}^{(\ell)}\right)^T,
\end{align}
where $h_u^{{t'}^{(\ell)}}$ is the node embedding of vertex $u$ at time $t'$ and after the $\ell$-th GCN layer, and $\mathbf{r}^{t^{(\ell})}$ and $\mathbf{Q}^{t^{(\ell)}}$ are trainable weights. Next, we use a GRU cell~\cite{GRU} and update node embeddings:
\begin{equation}
    {\zeta^{t}_u}^{(\ell)} = \mathrm{GRU}\left( h_u^{{t}^{(\ell)}}, \mathrm{short}_u^{t^{(\ell + 1)}}\right).
\end{equation}

\noindent
Note that the formulation of the \textit{Update Module} for \qencoder{} is the same as above, while we replace $h_u^{{t'}^{(\ell)}}$ with ${h_u}_Q^{{t'}^{(\ell)}}$. In this case, we denote the output as ${\zeta^{t}_u}_Q^{(\ell)}$. This process is illustrated in Figure~\ref{fig:model_framework}(b).

\head{FNN Layer}
Finally, a feedforward neural network is used to classify nodes based on the concatenation of the obtained embedding from the previous part:
\begin{equation}
    \psi^t_u = \mathrm{FNN}\left( {\zeta^{t}_u}^{(L)} \: \mathbin\Vert \:\: {\zeta^{t}_u}_Q^{(L)}\right),
\end{equation}
where $L$ is the number of GCN layers.

\head{Loss Function}
As discussed, the CS problem is treated as a binary classification task, where the output of model $\M$ is $\psi^t_q \in \mathbb{R}^{|\V^t|}$, representing the probability of each vertex $u \in \V^t$ being a member of the community $\CC^t_q$. Ground-truth labels ${\mathbf{y}^t_q}_u$ are defined for each vertex $u$ based on the community structure of the query set $\V^t_q$, and Binary Cross Entropy (BCE) is used as the loss function. The query-dependent loss function $\mathcal{L}^t_q$ is defined as:

\begin{equation}
\mathcal{L}^t_q = \frac{1}{|\V^t|} \sum_{u \in \V^t} - \left( {\mathbf{y}^t_{q}}_u \log({\psi^t_q}_u) + (1 - {\mathbf{y}^t_{q}}_u) \log(1 - {\psi^t_q}_u) \right).
\end{equation}
\noindent
By minimizing $\mathcal{L}^t_q$, the model learns to detect the community structure of the network by classifying vertices into their respective communities.

\begin{remark}
    Communities in real-world networks are known to be dynamic, with nodes joining or leaving communities over time. The proposed formulation allows for the adaptation of the model in scenarios where the ground-truth communities evolve over time. Consequently, during the training phase, the model can learn to capture the dynamics of the evolving ground-truth communities.
\end{remark}

\subsection{Training and Online Query}

\head{Training} 
We begin with a given set of query vertex sets denoted as $\mathcal{Q}_{\text{train}} = \{ q_1, q_2, \dots, q_n \}$ along with their respective ground-truth communities $\CC_{\text{train}} = \{ \mathcal{C}_1, \dots, \mathcal{C}_n \}$. First, we encode all query inputs as one-hot vectors. Then, at each time $t$, we repeatedly feed a query $q$ from $\mathcal{Q}_{\text{train}}$ into a model $\M$. The output $\psi^t_q$ from $\M$ is used to compute the loss and gradients of the model parameters. The updated parameters are then used for the next iteration, where $\M$ receives another query $q' \in \mathcal{Q}_{\text{train}}$ as input. This process continues until all queries in $\mathcal{Q}_{\text{train}}$ have been processed. The overall loss function is the sum of all query-dependent loss functions:
\begin{equation}
    \mathcal{L}^t = \sum_{q \in \mathcal{Q}_{\text{train}}} \mathcal{L}^t_q
\end{equation}

\begin{algorithm}[t]
    \small
    \caption{Temporal Community Identification}
    \label{alg:TCI}
    \begin{algorithmic}[1]
        \Require{A snapshot of a temporal network $\GG^t = (\V^t, \E^t, \X^t)$, a query vertex set $\V_q$, a trained model $\M$, and a threshold $\eta$}
        \Ensure{A temporal community $\CC_q$}
        \State $Q, \CC_q \leftarrow \V_q$; 
        \State $\psi_q \leftarrow$ feed $\V_q$ into $\M$;
        \While{$Q$ is not empty}
            \State pick and remove a vertex $u$ from $Q$;
            \For{$v \in N^t(u)$ and ${\mathbf{\psi}_q}_v \geq \eta$}
                \State $Q \leftarrow Q \cup \{v\}$;
                \State $\CC_q \leftarrow \CC_q \cup \{v\}$;
            \EndFor
        \EndWhile
        \Return $\CC_q$
    \end{algorithmic}
\end{algorithm}

\vspace{-2ex}
\head{Online Query}
Next, we describe the process of utilizing the pre-trained model $\M$ at time $t$ to identify the community $\CC^t_q$ of an online query $\V^t_q$ without the need for re-training $\M$. First, we construct a representative one-hot vector for $\V^t_q$ and feed it to $\M$. The output of $\M$ is denoted as $\psi^t_q \in \mathbb{R}^{|\V^t|}$, where ${\psi^t_q}_u \in [0, 1]$ represents the probability of vertex $u$ belonging to $\CC^t_q$ at time $t$. The online query stage is presented in Algorithm~\ref{alg:TCI}, where a threshold $\eta \in [0, 1]$ is utilized to identify the vertices in $\CC^t_q$ as those with ${\psi^t_q}_u \geq \eta$. Since communities are known to be connected, we use Breadth-First Search (BFS) traversal starting from query vertices to ensure connectivity.

\head{Time Complexity}
In Algorithm~\ref{alg:TCI}, the time complexity of GCN inferring (line 2) depends on the number of layers and the architecture of GCNs. BFS traversal (lines 3-7) takes $\mathcal{O}(|\E^t|)$ time. Accordingly, the time complexity is dominated by GCN inferring (line 2).

\subsection{Interactive Community Search}
The majority of community search methods in static and dynamic graphs utilize a progressive approach to identify the community~\cite{community_search_survey}. Nonetheless, to obtain a high-quality outcome that meets the user's expectations, the user may need to adjust their query, choose representative features, and modify the size of the community, based on each presented result for a given set of queries. In this section, we propose a meta-learning framework, where each query submitted by the user is considered an input to the \textit{Query Encoder} in a snapshot of the underlying network. We view the identification of communities based on user queries as separate tasks. 

\noindent
In this work, we consider a scenario where each user query is treated as a new snapshot of the underlying network, which can be either temporal or static. We aim to improve the performance of the model and user satisfaction by allowing the user to provide feedback on the output community, label data, and query new nodes. To update the model based on the user's feedback, we propose a meta-learning framework that uses a meta-model $\M^{(meta)}$ as a good initialization for deriving specialized models for future unseen user queries.

\noindent
To achieve this, we draw inspiration from previous work~\cite{roland} and adopt the Reptile algorithm~\cite{reptile}. Specifically, for each user, we first initialize the model $\M$ using $\M^{(meta)}$ and fine-tune it using back-propagation. We then update the meta-model by computing the moving average of the trained model:
\begin{equation*}
    \M^{(meta)} = (1 - \alpha) \M^{(meta)} + \alpha \M,
\end{equation*}
where $\alpha \in [0, 1]$ is the smoothing factor. It is worth noting that always fine-tuning the previous model may not be optimal, as the dynamics of each user's query can be different. Therefore, our proposal of finding a meta-model $\M^{(meta)}$ as a good initialization can lead to better specialized models. The process is presented in Algorithm~\ref{alg:ICS} and is illustrated in Figure~\ref{fig:model_framework}(c).

\begin{algorithm}[t]
    \small
    \caption{Interactive Community Identification}
    \label{alg:ICS}
    \begin{algorithmic}[1]
        \Require{A (temporal) network $\GG = (\V, \E, \X)$, \#epochs, a meta-model $\M^{(meta)}$, smoothing factor $\alpha$, and a threshold $\eta$}
        \Ensure{A temporal community $\CC_q$, and updated meta-model $\M^{(meta)}$}
        \State $\M \leftarrow \M^{(meta)}$;
        \While{user query vertex set $\V_q$ is not empty}
            \State $\CC_q \leftarrow$ feed $\V_q$ into $\M$ and find a community; \Comment{Algorithm~\ref{alg:TCI}}   
            \State $y_q \leftarrow$ user provides feedback on $\CC_q$ (labels data);
            \For{epoch = $1, \dots,$ \#epochs}
                \State Update $\M$ via backprop based on $\V_q$, $y_q$, and $\GG^t$;
            \EndFor
            \State user terminates or inputs a new query vertex set $\V_q \leftarrow \V_q^{(new)}$;
         \EndWhile

        \State $\M^{(meta)} \leftarrow (1 - \alpha) \M^{(meta)} + \alpha \M$;\\
        \Return $\M^{(meta)}$
    \end{algorithmic}
\end{algorithm}

\section{Experiments}

\head{Datasets}
We evaluate the performance of our proposed model using five real-world networks with ground-truth communities \cite{brain_dataset, ground_truth_community, football, FirmCore, FirmTruss}, from diverse domains, including social, communication, co-authorship, and brain networks. The statistics of these networks can be found in Table~\ref{tab:datastat}. Notably, the ground-truth communities in the \textit{brain} dataset correspond to functional systems in the human brain, making it a valuable demonstration of the effectiveness of our approach in identifying such systems. We conduct experiments using $100$ query-community pairs for the smaller \textit{football} and \textit{brain} networks, and $350$ query-community pairs for the other networks. We divide the queries into training, validation, and test sets, with a ratio of 40\% for training, 30\% for validation, and 30\% for testing. We use these sets to train the model, perform early stopping and hyperparameter tuning, and evaluate the end-to-end performance of our model, respectively.

% \vspace{-2ex}
\head{Setup}
Our models are implemented in Python 3.7 with \textit{PyTorch} and \textit{torch-geometric} libraries. The encoders consist of two GCN layers and a GRU unit with $h$ neurons. We fine-tuned the optimal value of $h \in \{32, 64, 128, 256\}$. In the ablation study, the effect of varying $h$ on the quality is evaluated.

\noindent
The reported times for the training and inference are based on the experiments on a Linux machine with \textit{nvidia A4000} GPU with 16GB of memory. We train the model for 100 epochs with a learning rate of 0.01. We used ReLU as the activation function and a dropout rate of 0.5. When dealing with large datasets, we follow the approach outlined in~\cite{GNN_CS, GNN_CS1, CS-MLGCN} and choose nodes within 2-hops of query nodes as possible subgraphs for each query. The model is then trained on these subgraphs to predict communities. Our code repository (\url{https://github.com/joint-em/CS-TGN}) contains more information about the implementation of our framework.

\begin{center}
\begin{table} [tpb!]
 \caption{Network Statistics}
 \vspace{-1ex}
    \resizebox{0.40\textwidth}{!} {
\begin{tabular}{l | c c c c c }
 \toprule
  {Dataset} & {\textit{football}} & {\textit{brain}}  &  {\textit{email}} & {\textit{dblp}}  &{\textit{youtube}}   \\
 \midrule 
 \midrule
    $|\V|$            &  114   &   190  &  1005 &  472K  &  1.1M  \\
    $|\E|$            &   613  &  720   &  25.5K &  1M   &  2.9M  \\
    \# Snapshots       &   5  &   4  &  6 &  8  &    4  \\
    \# Communities     &   12  &   15  &  42 &  5000   &  5000   \\
 \bottomrule
\end{tabular}
}
 \label{tab:datastat}
\end{table}
\end{center}

\vspace*{-6ex}
\head{Baselines} 
We compare our \model{} with state-of-the-art temporal and static community search (CS) methods. Reliable~\cite{reliable} is a $k$-core-based approach in temporal networks that finds a weighted $k$-core containing query nodes. Truss~\cite{truss-triangle} is an approach based on $k$-truss structure in temporal networks. MLGCN~\cite{CS-MLGCN} is a learning-based CS method in multiplex networks that leverages an attention mechanism to incorporate the information of different relation types. FTCS~\cite{FirmTruss} finds the connected FirmTruss containing query nodes with the minimum diameter. In multiplex methods, each snapshot of the network is treated as one layer or view. Finally, we compare our approach with a state-of-the-art learning-based method in static graphs, QD-GNN~\cite{GNN_CS}. In the interactive setting, we compare our method with ICS-GNN~\cite{GNN_CS1}, a GNN-based interactive community search model that requires re-training for each query.

\head{Queries and Evaluation Metrics}
The performance of all algorithms is evaluated using varying numbers of query nodes. We generated the queries randomly with a size between 1 to 10. The quality of the found communities $C$ is assessed through their F1-score, which measures the alignment with the ground truth $\tilde{C}$. The F1 score is defined as $F1(C, \tilde{C}) = \frac{2pre(C, \tilde{C})rec(C, \tilde{C})}{pre(C, \tilde{C}) + rec(C, \tilde{C})}$,
where $pre(C, \tilde{C}) = \frac{|C \cap \tilde{C}|}{|C|}$ and $rec(C, \tilde{C}) = \frac{|C \cap \tilde{C}|}{|\tilde{C}|}$. \eat{The running time is used to evaluate efficiency, with a cap of 5 hours and a memory footprint cap of 20 GB for all results. For index-based methods, the construction time is capped at 24 hours. }

\begin{figure}
    \centering
    \hspace{-3ex}
    \includegraphics[width=0.48\textwidth]{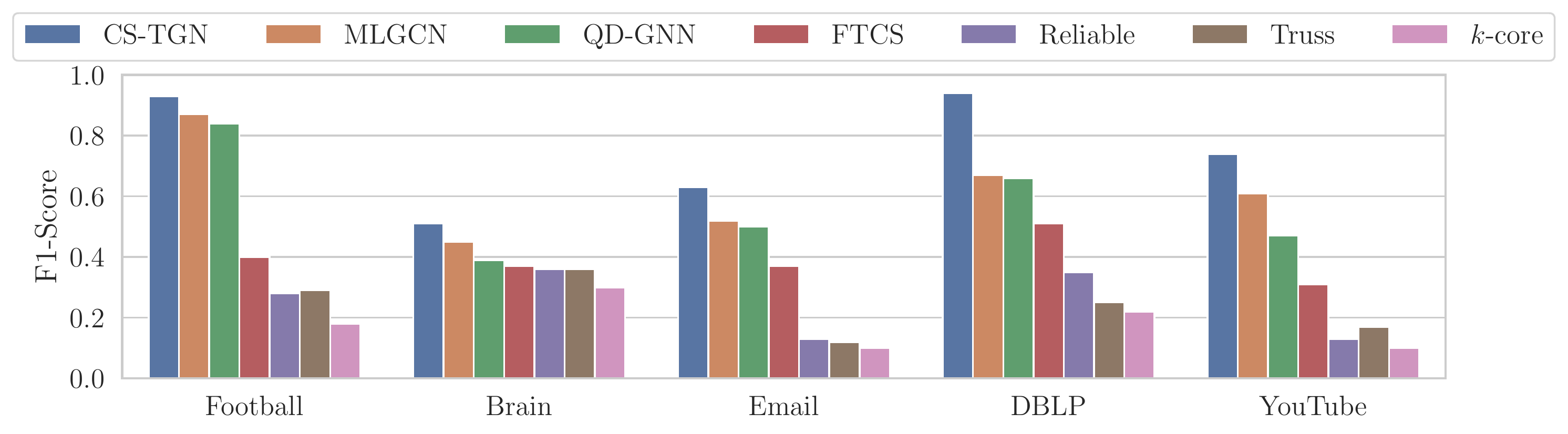}
    \vspace{-2.5ex}
    \caption{Quality evaluation.}
    \vspace{-1ex}
    \label{fig:F1-score}
\end{figure}

\head{Quality Evaluation}
The average F1-scores of our \model{} and the baselines on datasets with the ground-truth community are presented in Figure~\ref{fig:F1-score}. Our \model{} outperforms all baselines on all networks, with an average improvement of 12.6\%. This can be attributed to two factors. Firstly, the flexible nature of community structures is not well-captured by approaches that rely on pre-defined subgraph patterns, such as Reliable, Truss, $k$-core, and FTCS, resulting in inaccuracies for communities lacking such patterns. Secondly, while other learning-based methods neglect the temporal and historical properties of datasets, our approach can learn the community structure from the data and employs GRU cells with an attention mechanism to capture both short-term and long-term patterns in the network, thereby utilizing all the structural, temporal, and historical properties of the data.

\begin{figure}
    \centering
    \hspace{-3ex}
    \includegraphics[width=0.48\textwidth]{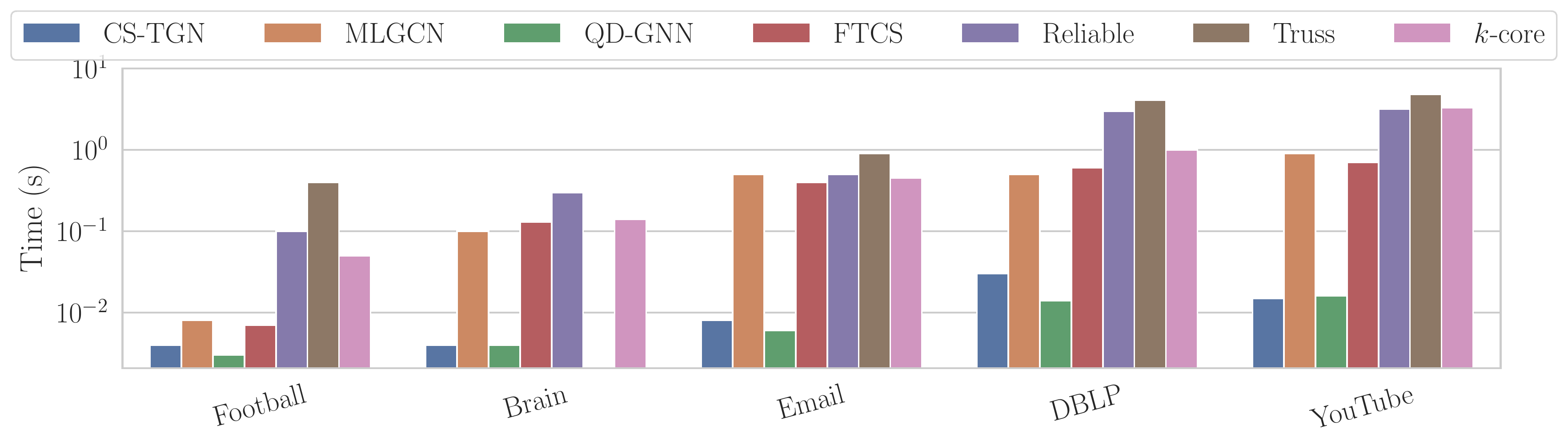}
    \vspace{-2.5ex}
    \caption{Query Efficiency.}
    \vspace{-1ex}
    \label{fig:Runing_time}
\end{figure}

\head{Query Efficiency}
  The efficiency of query processing for \model{} is evaluated in the test set. Figure~\ref{fig:Runing_time} presents the average query processing time of all methods for test queries. Our approach, \model{}, demonstrates a query time that is comparable to that of QD-GNN and surpasses other baselines in terms of efficiency. Notably, this finding is achieved by our method while also exhibiting superior performance compared to the baselines in quality evaluation.

\head{Effect of Temporal Learning} 
We conduct the following experiments to test our main hypothesis, that learning community structure over time is beneficial. First, we report the performance of the same architecture but without the GRU unit on the same networks and set of queries to quantify the quality difference and overhead caused by GRU units. Following that, we demonstrate how the number of snapshots affects the quality of results produced by the \model{} architecture.  

\noindent
Table \ref{tab:gru} presents a comparative analysis of the performance of \model{} against two of its variants that share the same architecture, albeit without the GRU cells. The first variant, referred to as the GRU-less version, is no longer temporal and is trained on the last snapshot of the network. The second variant, denoted as SUM-\model{}, employs a simple summation method to update node embeddings over time instead of using the GRU cell. In order to ensure a fair comparison, all test queries and answers are based on the last snapshot. Our experimental results demonstrate that \model{} outperforms both of its variants on all datasets in terms of F1-score, suggesting that capturing the community structure over time is beneficial for achieving superior performance.

\begin{figure}
    \centering
    \hspace*{-1ex}
    \includegraphics[width=0.16\textwidth]{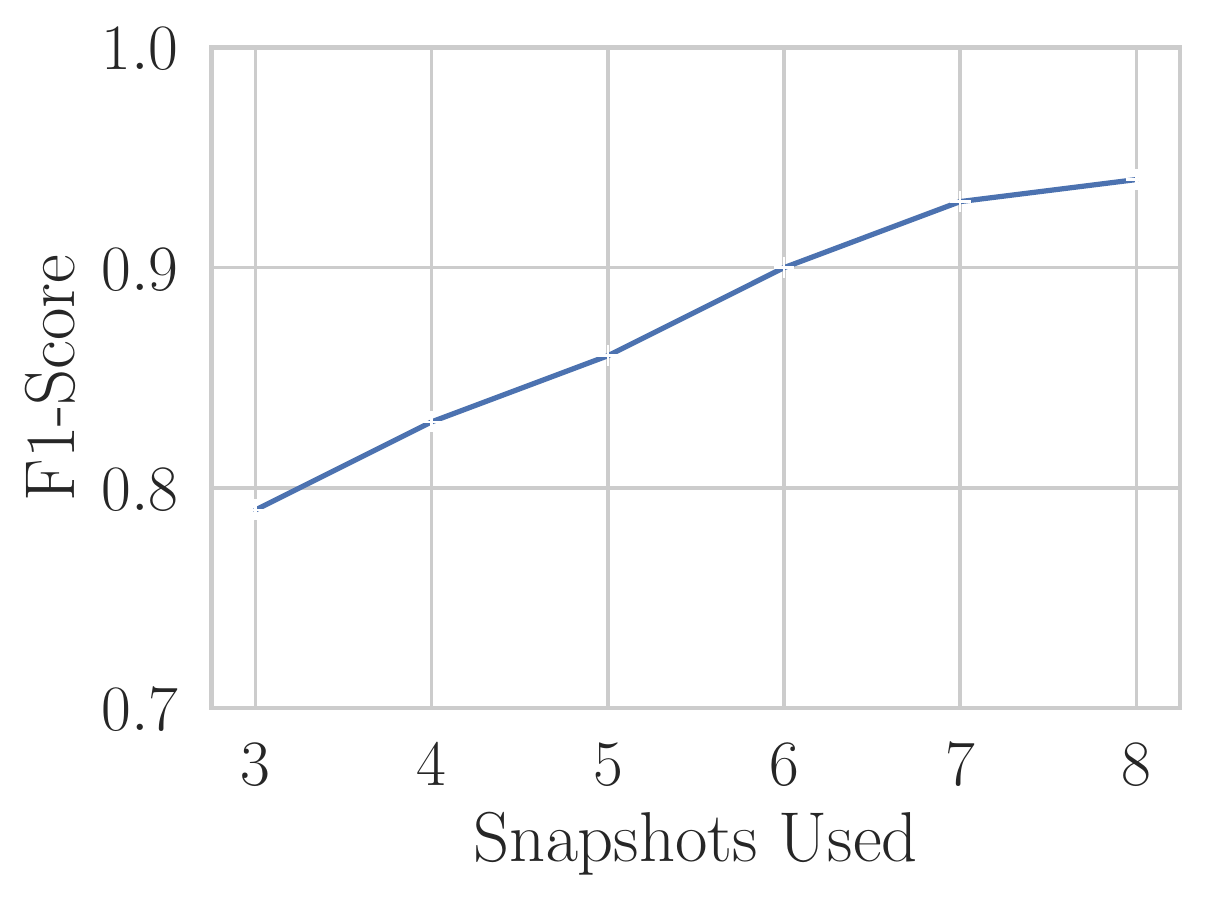}~
\includegraphics[width=0.16\textwidth]{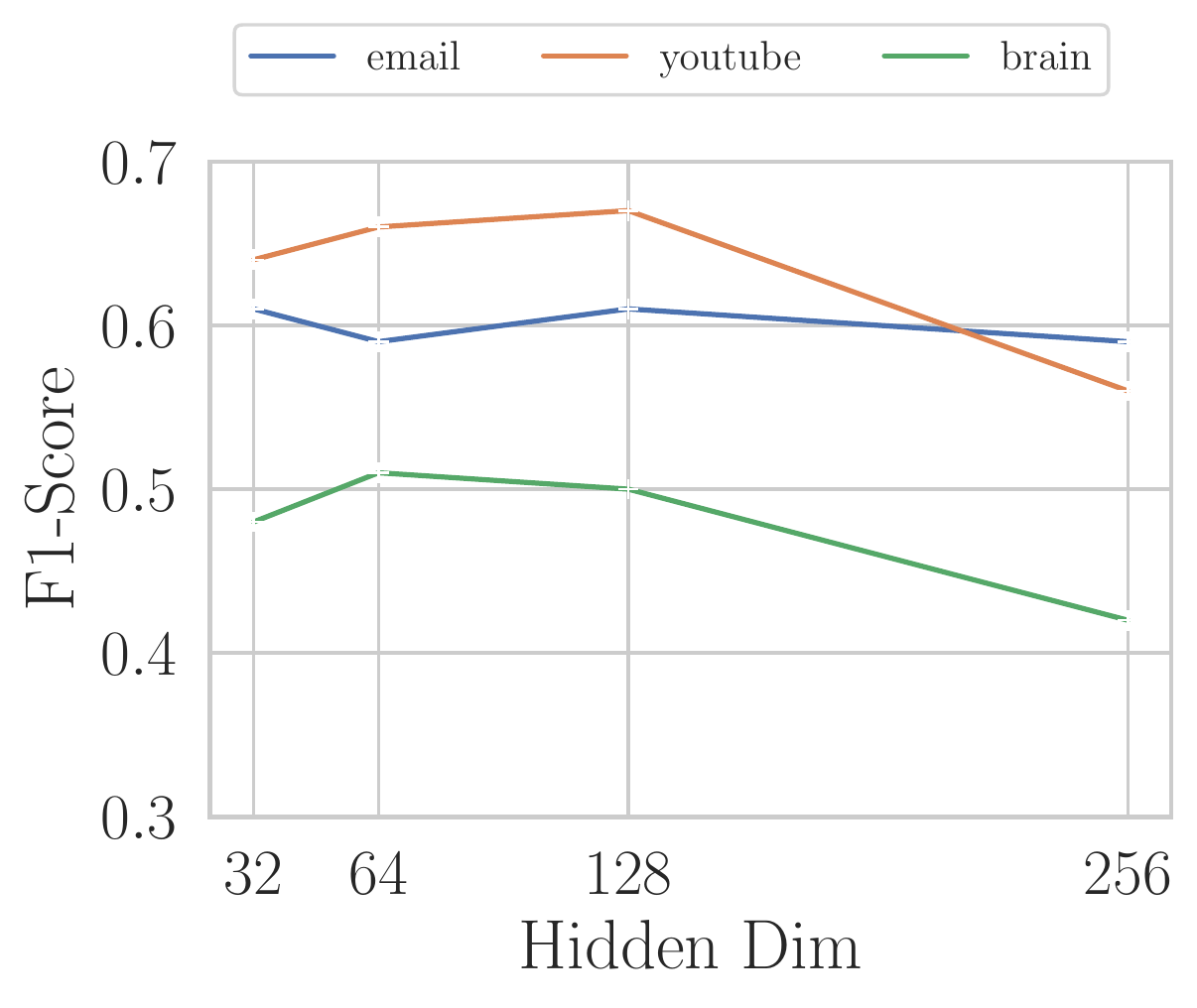}~
\includegraphics[width=0.16\textwidth]{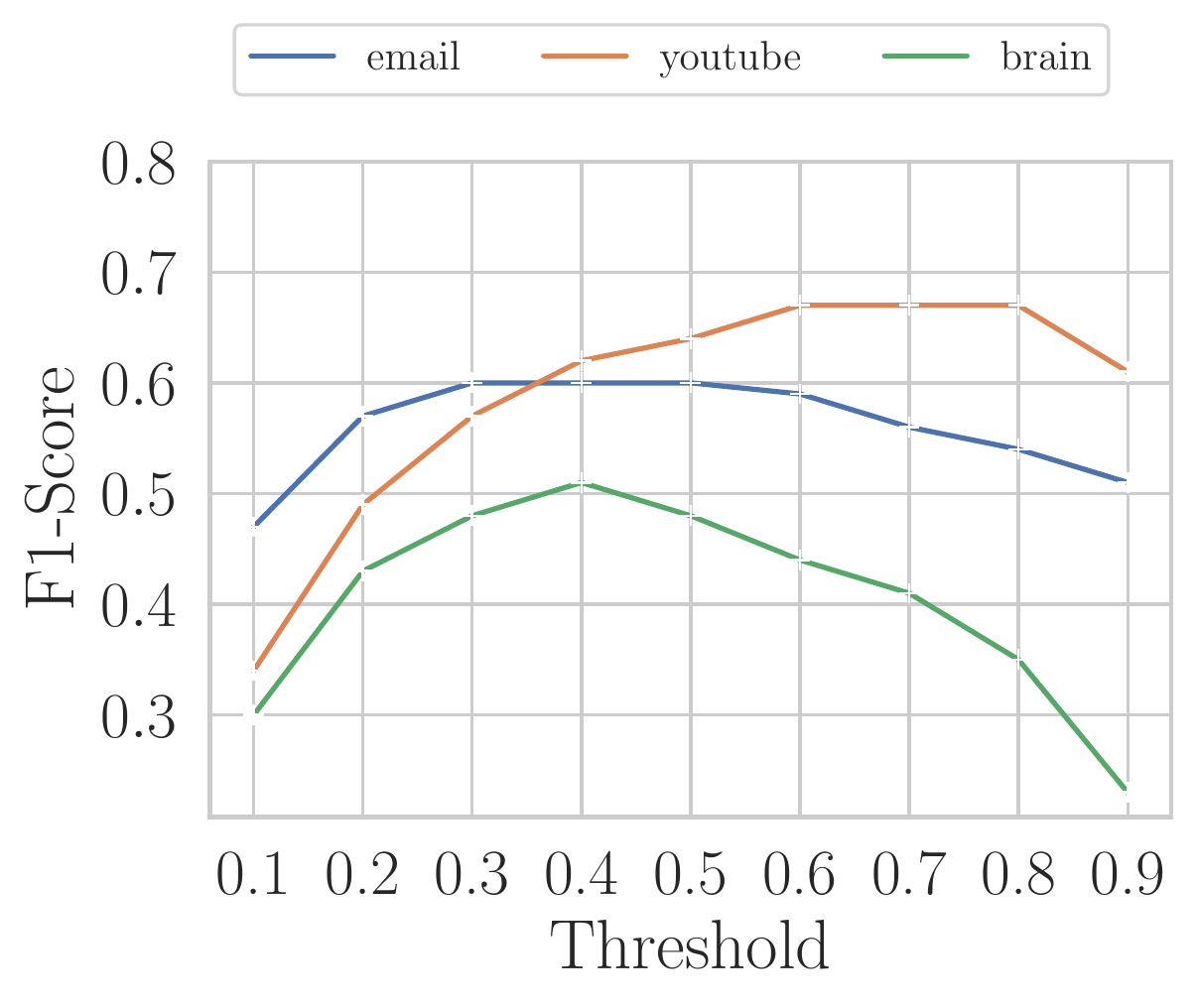}
    \caption{Effects of \#Snapshots (DBLP dataset) and Hyperparameters on F1-score.}
    \vspace{-2ex}
    \label{fig:abl}
\end{figure}

\noindent
To investigate the impact of the number of snapshots on the performance of \model{}, we conduct experiments by varying the number of snapshots on different versions of the DBLP dataset. As shown in Figure \ref{fig:abl}, increasing the number of snapshots leads to a significant improvement in performance. This result demonstrates the effectiveness of the attention-based update mechanism and the architecture of \model{}.

\begin{center}
\begin{table} [tpb!]
 \caption{The effect of GRU cells on F1-score}
 \vspace{-1ex}
    \resizebox{0.4\textwidth}{!} {
\begin{tabular}{l | c c c c c }
 \toprule
  {Dataset} & {\model} & {\model{} w/o GRU}  &  {SUM-\model{}}    \\
 \midrule 
 \midrule
    \textit{brain}    & \textbf{0.51} & 0.39            & 0.44               \\
    \textit{football} & \textbf{0.93} & 0.84            & 0.89               \\
    \textit{email}    & \textbf{0.63} & 0.50            & 0.51               \\
    \textit{youtube}  & \textbf{0.74} & 0.47            & 0.57               \\
    \textit{dblp}              & \textbf{0.94 }         & 0.66            & 0.83               \\
     \bottomrule
\end{tabular}
}
 \label{tab:gru}
\end{table}
\end{center}

\vspace{-2ex}
\head{Effects of Hyperparameters} 
We evaluate the effect of hyperparameters on the quality of our proposed model: \textbf{(1)}Number of neurons in the hidden layer (hidden dimension of GCN and GRU units) \textbf{(2)} threshold $\eta$. We demonstrated the effect of these hyperparameters for three datasets– \textit{email}, \textit{youtube}, and \textit{brain}–in Figure~\ref{fig:abl}. Grid search techniques can be used to tune these hyperparameters according to the dataset. Based on these results, the optimal threshold value $\eta$ is between 0.4 and 0.6, and the optimal hidden dimension is correlated with network size. 

\head{Interactive Community Search on Brain Networks}
Detecting and monitoring functional systems in the human brain is an important and fundamental task in neuroscience \cite{functional_system_brain, functional_system_brain2}. A brain network is represented as a graph, where nodes correspond to the brain regions and edges depict co-activation between regions. To identify the functional systems of each brain region, a community search method can be used. However, health-related domains are sensitive and require expert supervision. In this experiment, we apply two interactive community search methods on the \textit{brain} datasets while varying the size of the query vertex set. The results are presented in Table~\ref{tab:Intractive}. Our proposed \model{} outperforms the ICS-GNN model at any stage and number of interactions, thereby highlighting its effectiveness for interactive community search.

\begin{center}
\begin{table} [tpb!]
 \caption{Interactive community search on brain networks}
 \vspace{-1ex}
    \resizebox{0.45\textwidth}{!} {
\begin{tabular}{l | c c c c c c c c }
 \toprule
  {\#interactions} & {1} & {2}  &  {3} & {4} & {5} & {6} & {7} & {8}    \\
 \midrule 
 \midrule
    \model    & \textbf{0.33} & \textbf{0.35} & \textbf{0.39} & \textbf{0.41} & \textbf{0.38} & \textbf{0.47} & \textbf{0.55} & \textbf{0.56}   \\
    ICS-GNN & 0.24 & 0.27 & 0.29 & 0.35 & 0.31 & 0.36 & 0.4 & 0.42   \\
     \bottomrule
\end{tabular}
}
 \label{tab:Intractive}
\end{table}
\end{center}

\section{Conclusion}
In this paper, we propose a novel approach for community search in temporal networks called \model{}. Our approach is a query-driven temporal graph convolutional neural network that takes a data-driven approach to capture flexible community structures and incorporates dynamic temporal properties from ground-truth communities. To achieve this, \model{} employs two encoders to encode the local query-dependent structure and global query-independent graph structure. To capture both short-term and long-term patterns and update node embeddings over time, \model{} uses a contextual attention mechanism and GRU cells. Additionally, we show how \model{} can be used for interactive community search and formulate the problem as a meta-learning approach over temporal networks. We evaluate our model on several real-world datasets with ground-truth communities and demonstrated its superior performance compared to existing state-of-the-art methods in terms of accuracy and efficiency.

\bibliographystyle{ACM-Reference-Format}
\bibliography{main}

% \appendix

% % \section{Appendix}

\end{document}